\documentclass[12pt, a4paper, twocolumn]{article}
\usepackage{graphicx} 
\usepackage{authblk}
\usepackage{booktabs}
\usepackage{caption}
\usepackage{parskip}
\usepackage{sectsty}
\allsectionsfont{\centering}
\usepackage{hyperref}
\usepackage[english]{babel}
\usepackage{mathptmx}
\usepackage[backend=biber, style=numeric, sorting=none, doi=true, url=false, isbn=true]{biblatex}
\usepackage[margin={2cm,2cm}]{geometry}
\title{\textbf{In situ Imaging of Precipitate Formation in Additively Manufactured Al-Alloys by Scanning X-ray Fluorescence}}
\date{}
\author[1]{Isac Lazar\footnote{Corresponding author. Contact: isac.lazar@sljus.lu.se}} 
\author[2]{Bharat Mehta}
\author[3]{Vendulka Bertschová}
\author[2]{Sri Bala Aditya Malladi}
\author[4]{Zhe Ren}
\author[5]{Srashtasrita Das}
\author[4]{Johannes Hagemann}
\author[4]{Gerald Falkenberg}
\author[2]{Karin Frisk}
\author[1]{Anders Mikkelsen}
\author[2]{Lars Nyborg}
\affil[1]{\small Division of Synchrotron Radiation Research, Department of Physics, and NanoLund, Lund University, 221 00 Lund, Sweden}
\affil[2]{Department of Industrial and Materials Science, Chalmers University of Technology, Rännvägen 2A, Göteborg, 41296 Sweden }
\affil[3]{Tescan Orsay Holding a.s, 623 00 Brno, Czech Republic}
\affil[4]{Deutsches Elektronen-Synchrotron DESY, Notkestr. 85, 22607 Hamburg, Germany}
\affil[5]{Institute of Chemical Technology and Polymer Chemistry (ITCP), Karlsruhe Institute of Technology (KIT), 76131 Karlsruhe, Germany}
\begin{document}
\maketitle
\textbf{\textit{Abstract ---}
A new family of high-strength Al-alloys has recently been developed, tailored for the powder bed fusion-laser beam process. In these alloys, Mn, Cr and Zr are incorporated in solid solution at amounts up to three times that of equilibrium in the as-printed state.  Mn and Cr-enriched precipitates that form during printing and heat treatment influence the material's mechanical properties. In this study, direct imaging of these precipitates was accomplished through the utilisation of in situ synchrotron-based scanning X-ray fluorescence. During heat treatment, a selective accumulation of Cr and Mn in two distinct types of precipitates at grain boundaries was observed. Additionally, the microstructure at the melt-pool boundary, containing precipitates found in the as-printed state, remains thermally stable during the heat treatment. The study demonstrates the significant value of employing high-sensitivity in-situ X-ray fluorescence microscopy in exploring the kinetics of sub-micrometre scale precipitation.  }
\section*{Introduction}
Powder bed fusion-laser beam (PBF-LB) is an additive manufacturing (AM) process that leverages selective melting of a thin layer of metal powder (20-80 µm thick) using a laser of small spot diameter (40-100 µm) \cite{ref1}\cite{ref2}. The process entails a high cooling rate (10$^3$-10$^5$ K/s) which for aluminium (Al) alloys enables higher supersaturation of certain alloying elements \cite{ref3}\cite{ref4}. Using rapid solidification, previously more than 10 wt\% of Mn and Cr have been dissolved in solid solution in Al-alloys. These alloys were then strengthened via precipitation hardening \cite{ref5}\cite{ref6}\cite{ref7}\cite{ref8}, a process that is widely acknowledged as the most efficient mechanism for increasing strength in Al-alloys \cite{ref9}\cite{ref10}\cite{ref11}. As-printed Al-alloys produced through the PBF-LB process demonstrate similar supersaturation of solutes and quenched-in vacancies \cite{ref12}\cite{ref13}\cite{ref14} thus enabling the exploration of novel precipitation reactions during heat treatment of additively manufactured parts. Recently, we have developed and investigated a new type of Al-Mn-Cr-Zr alloy specifically tailored for the PBF-LB process. It was shown that in addition to high strength in the as-printed state, precipitation hardening by direct ageing results in a 40 Vicker’s Hardness (HV) increase reaching 142 HV in peak aged condition \cite{ref15}. Based on ex-situ studies \cite{ref14}\cite{ref15}, the hardness contribution of Mn and Cr is thought to arise from solid solution strengthening and the formation of needle-/ plate-like Mn-rich precipitates during heat treatment. In addition, the microstructure has been found to contain Mn-rich precipitates at melt pool boundaries and along solidification boundaries in as-printed conditions. After heat treatment, large Mn-rich precipitates were also formed at grain boundaries, which can potentially compromise ductility. Finally, Cr has been detected inside many of the above-mentioned precipitates, however, its exact behaviour during printing and heat treatment remains uncertain. In this study, our investigation focuses on the distribution of Cr and Mn at grain boundaries and melt pool boundaries during heat treatment of the recently developed Al-Mn-Cr-Zr alloy. 

Other authors have investigated solid-liquid interphases \cite{ref16} or precipitation behaviour in as-atomised Al-Cr-Mn-Co-Zr alloy powders and hot extruded samples using in-situ heat treatment in the STEM \cite{ref17}\cite{ref18}. These studies depicted the precipitation of a phase rich in Mn and Cr. However, no elemental mapping during heating was reported. In the present study, we use synchrotron scanning X-ray fluorescence (XRF) for in-situ elemental mapping during heat treatment of the as-printed alloy. Using scanning XRF it is possible to spatially resolve the distribution of elements even at very low concentrations \cite{ref19}. Using this technique, we gain further insight into the microstructure of the material and its evolution during heat treatment.
\section*{Materials and Methods}
\begin{figure*}[h]
    \centering
    \includegraphics[width=\linewidth]{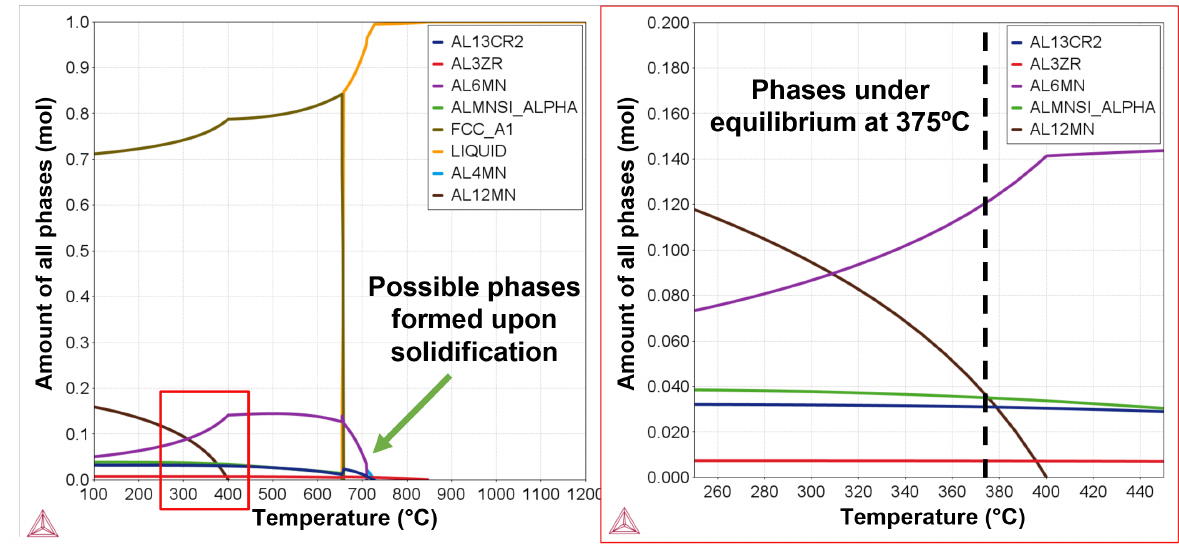}
    \caption{ThermoCalc one-axis simulation for alloy C which provides a calculated estimate of the stable phases at different temperatures. The inset in red shows phases that may form upon heat treatment at 375 ºC. Image redrawn similar to \cite{ref15}, calculations conducted with COST507 database \cite{ref20} using ThermoCalc 2022a}
    \label{fig1}
\end{figure*}
\begin{figure*}[h]
    \centering
    \includegraphics[width=0.9\linewidth]{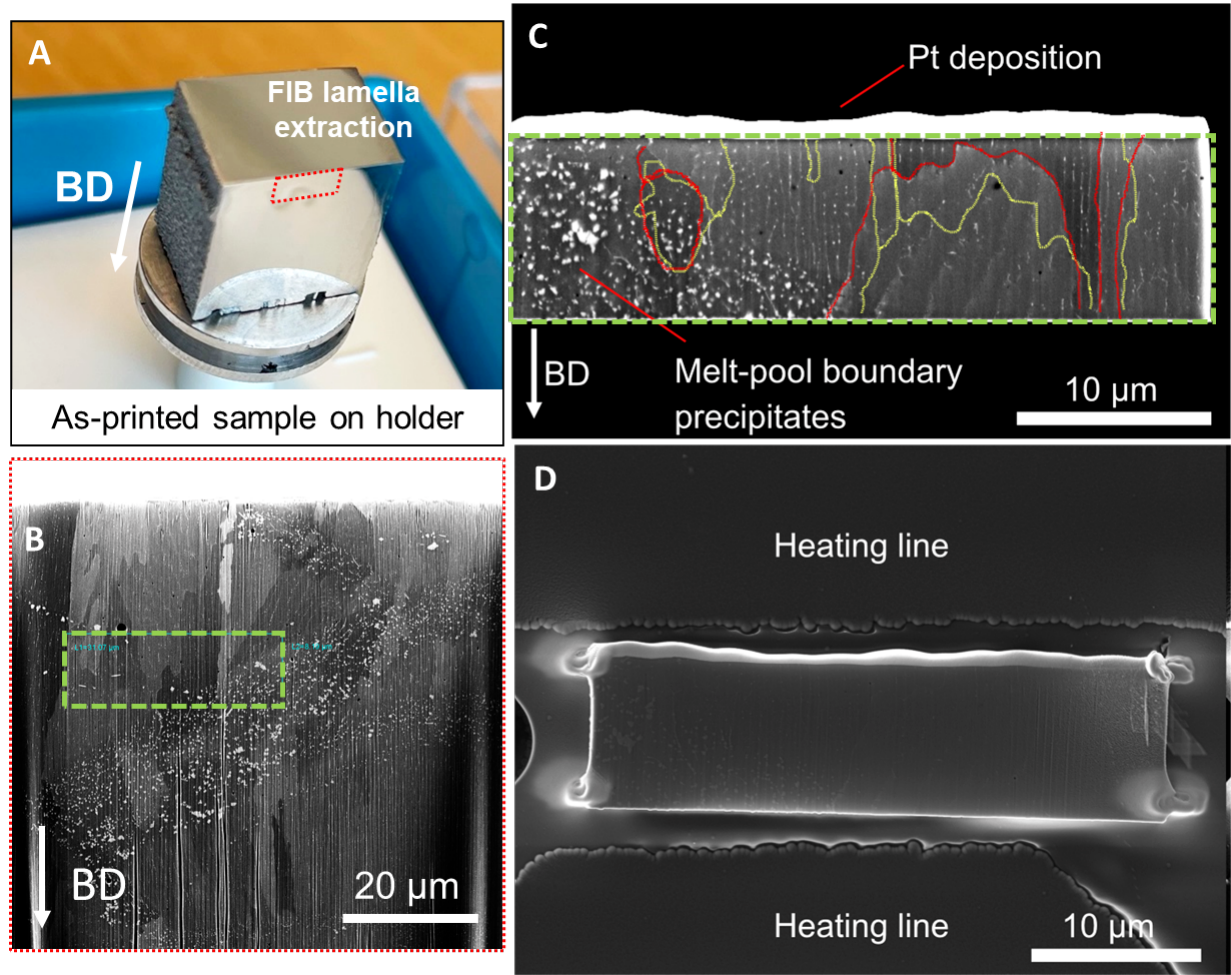}
    \caption{Lamella preparation. A) shows the as-printed sample showing the location of FIB lamella extraction with build direction (BD) marked. B) is a BSE image of the sample in FIB-SEM wherein the location for final lamella extraction is marked in green. C) shows the thinned-down lamella before placement on the heating chip. The bright layer on top of the lamella is a protection Pt deposition. Red and yellow lines mark the location of grain boundaries, as seen from both sides of the lamella. D) Shows the placement of the lamella between two heating lines on the in-situ heating chip. Corners are glued by Pt deposition.}
    \label{fig2}
\end{figure*}
\begin{center}
\captionof{table}{Chemical composition (in wt\%) of alloy C in as-printed condition. } 
\begin{tabular}{@{}lllllll@{}}
\toprule
\textbf{Element} & \textbf{Al} & \textbf{Mn} & \textbf{Cr} & \textbf{Zr} & \textbf{Fe} & \textbf{Si} \\ \midrule
wt\%             & Bal.        & 5.0         & 0.8         & 0.6         & 0.16        & 0.17       
\end{tabular}%
\label{tab:my_table}
\end{center}
Cubic samples (10 mm side) with $>$99.8\% relative density were manufactured using an EOS M290 PBF-LB machine equipped with a Yb-fiber laser. Parameters used for printing were 370 W power, 1300 mm/s laser speed, 0.13 mm hatch distance and 0.03 mm layer thickness \cite{ref21}. A standard scan rotation of 67° was applied between each layer. The samples were made from as-atomised powder (20-53 µm) of Alloy C, see Table \ref{tab:my_table}. Fe and Si are impurities from the atomisation process \cite{ref15}. A one-axis equilibrium curve drawn using ThermoCalc is shown in Figure \ref{fig1} as a guide to understanding the possible phases formed in the material upon solidification and during heat treatments at 375 ºC. 

A lamella of material in as-printed condition was extracted and thinned using a Xe$^+$ Plasma Focused Ion Beam (PFIB) SEM (TESCAN AMBER X microscope) (Figures \ref{fig2}B)-D)). The inert Xe$^+$ ion source PFIB was used to avoid Ga contamination of grain boundaries, commonly occurring with liquid Ga FIB systems \cite{ref22}\cite{ref23}\cite{ref24}. Rocking stage polishing was utilised as the final preparation step, mitigating curtaining artefacts on the lamella surfaces.  The lamella was roughly 10 µm and 30 µm in height and width respectively, and 1 µm in thickness. The final lamella can be seen in Figure \ref{fig2}C). Extraction from the cubic sample (Figure \ref{fig2}A) was done along the build direction, containing regions with both melt pool boundary precipitates and solidification boundary precipitates \cite{ref14}. The lamella was glued onto a Climate in-situ heating chip (DENSSolutions) by Pt deposition in the corners as seen in Figure \ref{fig2}D).

In-situ scanning X-ray fluorescence measurements were performed under N$_2$ atmosphere using the Climate (DENSsolutions) sample holder with a designed interface adapted for hard X-ray microscopy at the P06 synchrotron beamline at PETRA III, DESY, Hamburg \cite{ref25}\cite{ref26}. The flux was around 10$^8$ ph/s with a photon energy of 9 keV. The spot size on the sample was estimated to be 70 $\times$ 70 nm$^2$ from ptychographic reconstructions of the probe. Fluorescence was collected using an annular detector (Rococo 3, PNDetector), placed directly upstream from the sample centred around the beam. Fluorescence data was normalised to the incoming beam intensity and fitted with the pyMCA package \cite{ref27}. Images were registered using the OpenCV python package. 

In the experiment, the elemental changes due to annealing at 375 ºC were to be followed in detail, and thus the following procedure was adopted. For capturing microstructural evolution on different length scales, as well as high temporal and spatial resolutions, four different types of scans were performed. Three shorter in-situ scans were conducted while the sample was held at 375 ºC. The first scan was done on the complete lamella with a step size of 150 nm and an exposure time of 20 ms (not shown) for the capture and correction of sample drift. Two regions of interest spanning 5 µm $\times$ 5 µm were then scanned using a 50 nm step size and 120 ms exposure time. One region of interest was at a grain boundary and the second at a melt pool boundary as illustrated in Figure \ref{fig3}. These three scans were performed in a sequence once every hour. At hours 0, 4, 8, 14, 16, and 24, the sample was cooled to room temperature, where longer scans were performed of the whole lamella, with 50 nm step size and 30 ms exposure time. Cooling of the sample was to prevent significant microstructural changes during the longer, 90-minute scan. Heating and cooling rates were 15-20 ºC/min and 45-50 ºC/min respectively to avoid sudden contraction or expansion of the lamella. It was previously shown that no microstructural changes occur below 300 ºC \cite{ref15}, thus the heating and cooling lead to minimal microstructural change ($<$5 min per cycle). Additionally, comparing this approach to continuous ex-situ heat treatment showed consistent qualitative microstructural features (see appendix figure \ref{figA1}) thus validating our methodology to represent precipitation in-situ for this alloy system.
\section*{Results}
\begin{figure*}
    \centering
    \includegraphics[width=\linewidth]{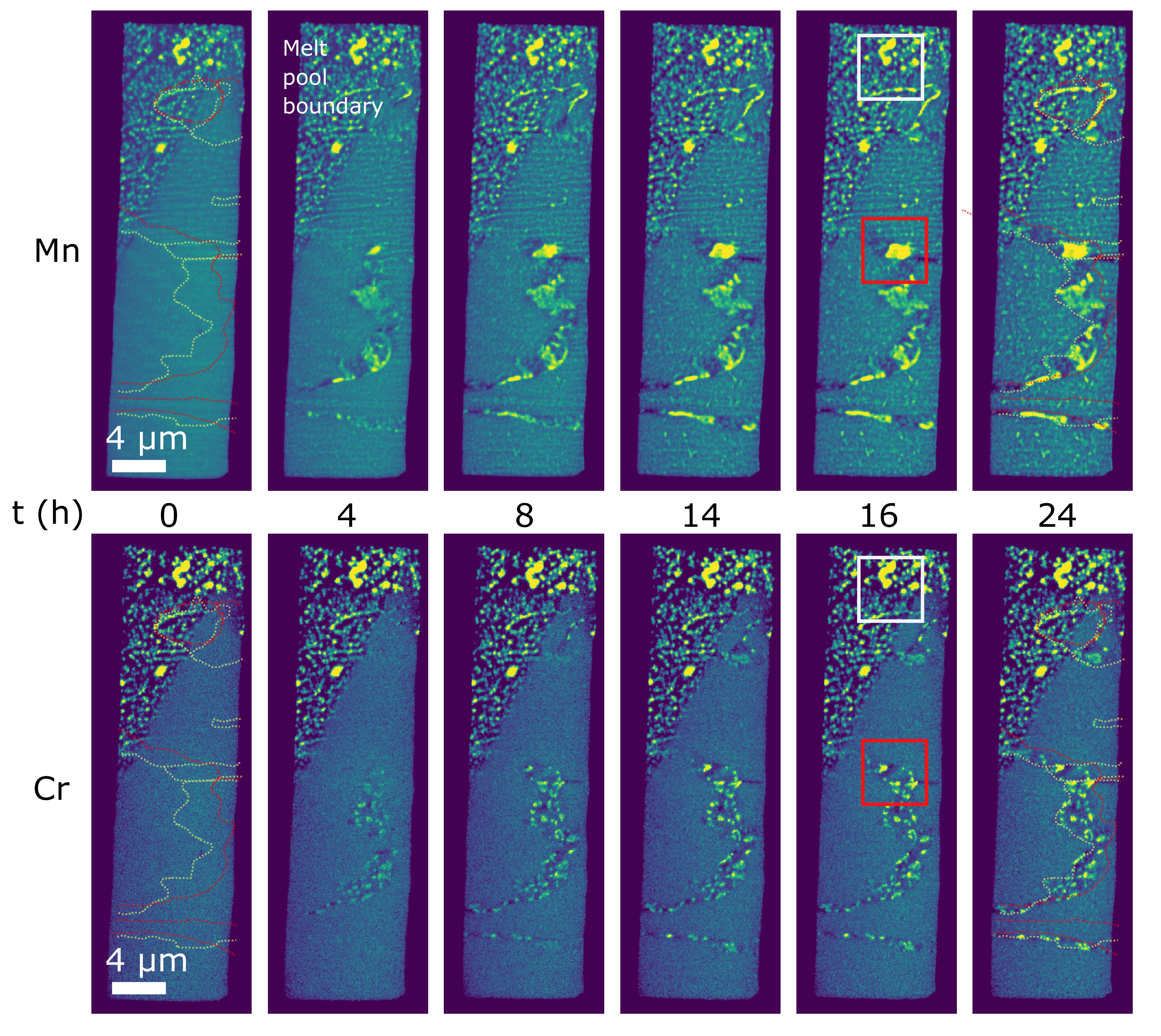}
    \caption{Microstructural evolution during 24 hours of heat treatment at 375 °C illustrated by the elemental distribution of Mn (top) and Cr (bottom). Dashed red and yellow lines indicate grain boundaries as observed from each opposite face of the lamella. Over time, Mn and Cr enrichment occurs in grain boundary regions, accompanied by slower growth of smaller precipitates inside grains. The regions highlighted in red and white squares are further presented in Figures \ref{fig4} and \ref{fig5} respectively, containing grain boundary precipitates and melt pool boundary precipitates.}
    \label{fig3}
\end{figure*}
In Figure \ref{fig3}, the Mn and Cr distribution in the whole lamella can be seen, as a function of heat treatment time. Before heating (time 0 h), precipitates containing Mn and Cr can be seen in the melt pool boundary and a few small Mn-rich precipitates in solidification boundaries. The elemental distribution in the rest of the lamella appears homogeneous. Already after 4h, Mn- and Cr-rich precipitates can be seen growing in the lamella. Although the lamella is only 1 µm thick, it is important to note that the recorded images are the projections of the 3D elemental distributions through its thickness. Thus the precipitates can appear larger in projection, due to them forming along a tilted grain boundary. By noting the grain boundary structure using SEM images on both sides of the lamella before placement on the chip, it was possible to construct regions wherein grain boundaries were present, assuming they extend linearly through the depth of the lamella. The grain boundaries are constrained as illustrated by the coloured dashed lines at times 0 h and 24 h in Figure \ref{fig3}. This shows that both Mn and Cr enrichment occurs at grain boundaries, confirming results from previous, ex-situ studies \cite{ref15}. Additionally, slower growth of numerous small precipitates inside grains occurs, as can be seen in the Mn elemental distribution at 14 h–24 h.

\begin{figure*}
    \centering
    \includegraphics[width=\linewidth]{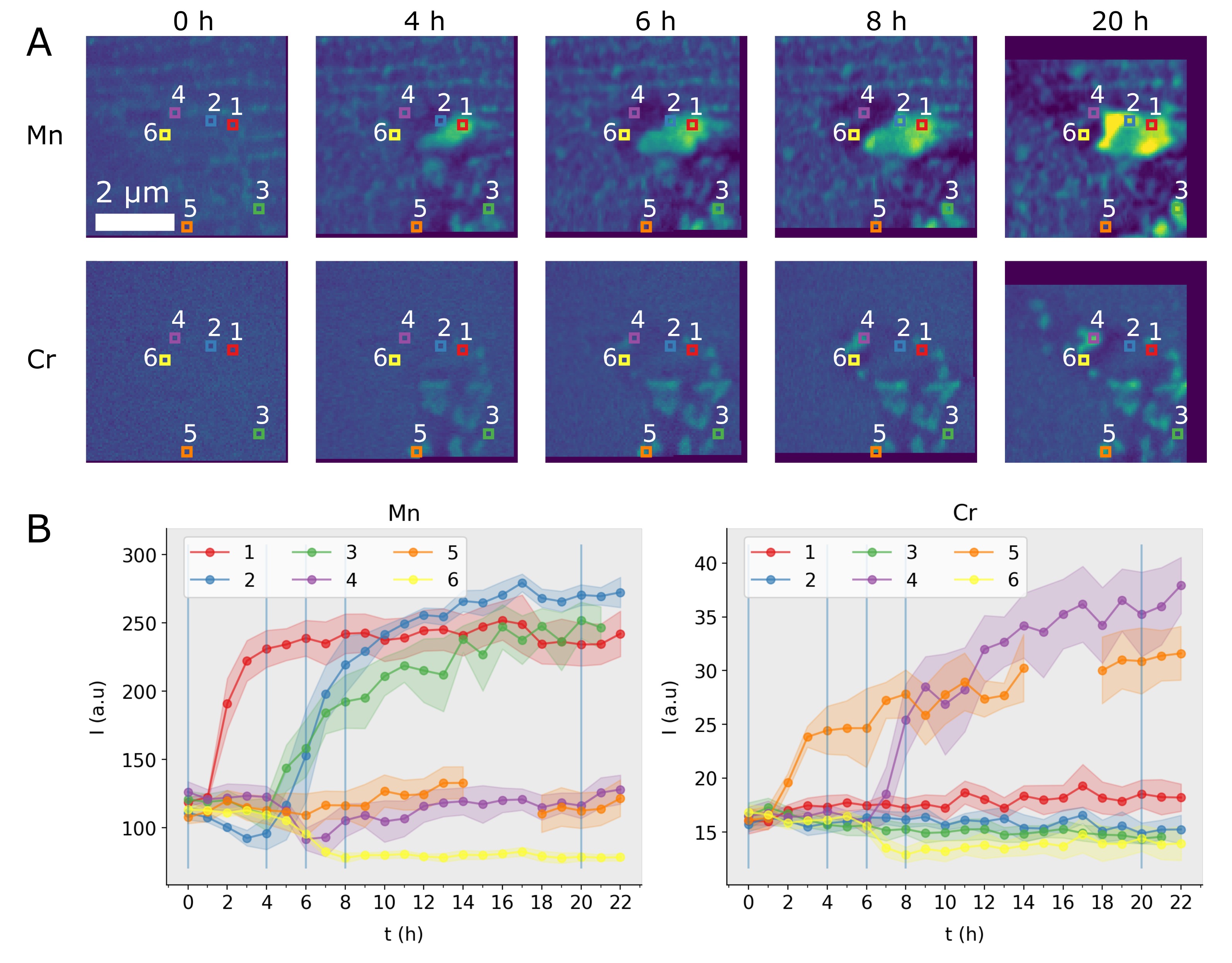}
    \caption{Evolution of Mn and Cr at grain boundary region, highlighted by the red square in Figure \ref{fig3}. A) Shows the elemental distribution at selected snapshots of heat treatment at 375 °C, for Mn (top row) and Cr (bottom row). B) shows the Mn and Cr intensity respectively in 6 locations defined by colored squares in A). Vertical blue lines mark the time snapshots seen in A). Shaded areas indicate the uncertainty in quantified intensity. Location 5 has a broken line due to this location being outside the imaged region in some frames due to sample drift}
    \label{fig4}
\end{figure*}
\begin{figure*}
    \centering
    \includegraphics[width=\linewidth]{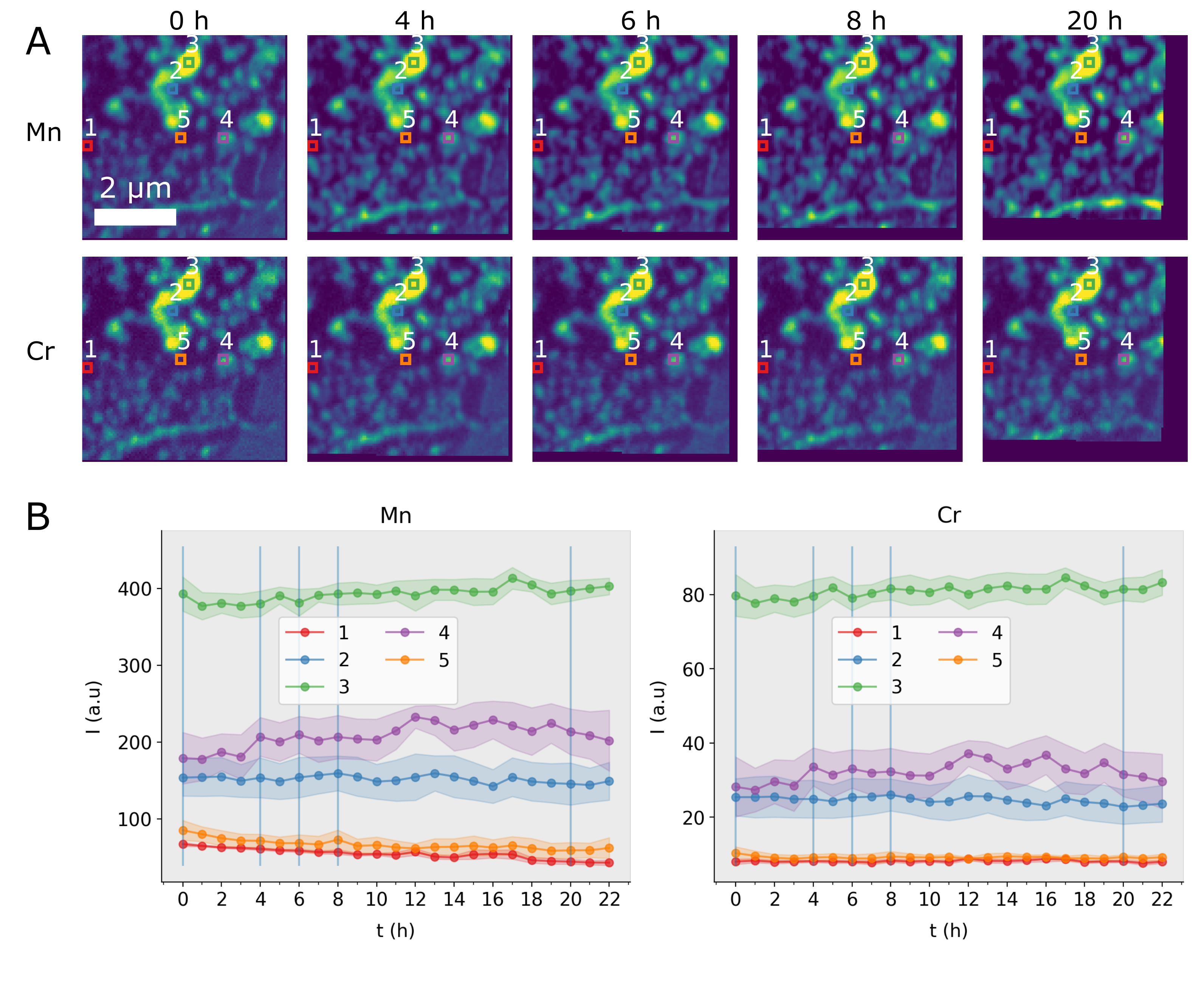}
    \caption{Evolution of Mn and Cr at melt pool boundary region, as highlighted by the white square in Figure \ref{fig3}. A) Shows the elemental distribution at selected snapshots of heat treatment at 375 °C, for Mn (top row) and Cr (bottom row). The melt pool boundary precipitates contain both Mn and Cr and are present in the as-printed condition. B) shows the Mn and Cr intensity respectively in 5 locations defined by colored squares in A). Vertical grey lines mark the time snapshots seen in A). Shaded areas indicate the uncertainty in quantified intensity. }
    \label{fig5}
\end{figure*}
In Figure \ref{fig4}4A) the details of the microstructural evolution at a grain boundary can be seen with high temporal and spatial resolution. In addition, at selected areas of interest (indicated by numbers 1 to 6) the evolution of Mn and Cr fluorescence intensities over time has been obtained and is shown in Figure \ref{fig4}B). Similar to the overview results, the initial distribution of elements is quite homogeneous. During heat treatment, Mn forms precipitates accompanied by depletion of Mn in the surrounding matrix. In Figure \ref{fig4}A), a noticeable depletion of Mn in the vicinity of a growing precipitate can be observed. This depletion becomes even more apparent when examining the Mn intensities in Figure \ref{fig4}B), particularly for locations 1, 2, and 3. Location 1 is where one precipitate forms which sees a rapid increase in Mn intensity after only 1 hour, before saturating after 6 hours. In location 2, the Mn intensity initially decreases, due to the depletion of Mn, after which it increases when the precipitate grows reaching the same saturated levels as in location 1. In location 6, the matrix is continuously depleted of Mn up until 8 hours, after which the concentration stays the same for the rest of the heat treatment time. Location 3 captures the growth of another Mn-rich precipitate, highlighting differences in the onset of growth. Interestingly Fe is also seen incorporated in these Mn-rich precipitates as evidenced by Figure \ref{figA2} in the appendix, which is in contrast to Cr. Cr enters into smaller precipitates surrounding the larger Mn precipitate. The earliest observable onset of Cr enrichment can be seen after 2 hours of heat treatments. In contrast to locations 1 and 2, at locations 4 and 5, the growth of Cr-rich precipitates is evident by the steep increase in Cr intensity. The minor decrease in Cr intensity in location 6 is due to the depletion of Cr, close to a larger growing Cr-rich precipitate (location 4). 

Figure \ref{fig5} shows the analysis of the microstructural evolution at the melt pool boundary region. The as-printed precipitates present in this region are enriched in both Mn and Cr, with a ratio that remains stable over time. The intensity in some select locations is highlighted by the plots in Figure \ref{fig5}B) illuminating the limited activity in this region. A slight matrix depletion may be present in regions in between precipitates at locations 1 and 5. This may be due to a supply of Mn to the growing grain boundary precipitate in the lower right corner of the imaged region. Compared to the trends seen for Cr in Figure \ref{fig4}B), any enrichment or depletion of Cr in Figure \ref{fig5}B) is difficult to discern.
\section*{Discussion}
The use of in-situ scanning XRF is proven to be a useful tool for observing precipitation behaviour mimicking bulk precipitation, thus providing useful insight into this new material family. In particular, the ability to study relatively thick samples yielding information closer to what is found in the bulk, combined with the $<$100 nm resolution and the opportunity to distinguish and quantify several elements in the material can be emphasized.  Still, the combination with other ex-situ and in-situ imaging techniques is needed as they in combination can give a detailed picture of structure and kinetics as well as allow the exclusion of experimental artifacts \cite{ref14}\cite{ref15}\cite{ref17}\cite{ref18}\cite{ref19}. 

In this study, we see the formation of an Mn-rich phase at grain boundaries, previously identified in ex-situ SEM studies of heat-treated samples as either the Al$_{12}$Mn or Al$_6$Mn phase, consistent with theoretical predictions depicted in Figure \ref{fig1} \cite{ref15}. This phase also seems to integrate Fe, a finding not directly observed before, likely obscured by Fe's low concentrations. Theoretical models propose that even small additions of Fe may enhance the stability of the Al$_6$Mn phase in the Al-Mn-Fe system, potentially serving as an indicator of this phase's development \cite{ref15}. This observation highlights the advantage of employing the XRF technique, which exhibits lower detection limits than that of conventional electron-beam-induced energy dispersive spectroscopy (EDS) \cite{ref19}. 

Co-precipitation of a distinct unknown Cr-rich phase occurs adjacent to the Mn-rich precipitates, a phenomenon that has not been seen before. Further studies are being conducted to classify this new phase, where viable structure candidates are the types Al$_6$Mn, Al$_{13}$Cr$_2$, Al$_{12}$Mn or alpha-AlMnSi, based on theoretical predictions in Figure \ref{fig1}.  

The present study also shows that the microstructure is thermally stable over time in the melt pool boundary. The precipitates found here are present in the as-printed material, and they do not grow significantly during heat treatments. These precipitates are enriched in both Mn and Cr, and no significant change in Mn or Cr distribution can be observed. They are likely of types Al$_6$Mn, Al$_{13}$Cr$_2$, or alpha-AlMnSi, according to theoretical predictions. 

Further investigation involving CALPHAD-based database development and TEM diffraction characterization of all the abovementioned precipitates will be part of future studies. The information gathered in this study will prove valuable for understanding the kinetics of precipitate formation in these alloys.
\section*{Conclusion}
In summary, this study focuses on experimentally examining the Mn and Cr precipitation kinetics in a novel Al-Mn-Cr-Zr based alloy, using in-situ scanning X-ray fluorescence. We observed the microstructural evolution in-situ during 24 hours of heating by the Mn and Cr elemental distributions. At grain boundaries, co-precipitation of separate Cr-rich and Mn-rich phases occurs. At the melt-pool boundary, the microstructure remains stable over time, retaining precipitates enriched in both Mn and Cr. Thus, Cr plays an active role in the precipitation kinetics of this new alloy.
\section*{Acknowledgements}
This work was supported by the German Federal Ministry of Education and Research (BMBF) project COSMIC (no. 05K19VK4). \newline
We acknowledge DESY (Hamburg, Germany), a member of the Helmholtz Association HGF, for the provision of experimental facilities. Beamtime was allocated for proposal I-20220364 EC. \newline
Support from Production Area of Advance at Chalmers University is acknowledged. \newline
Parts of this research were carried out at P06 PETRA III beamline, and we would like to thank Andreas Schropp, Dennis Brueckner, Mikhail Lyubomirskiy and Shweta Singh for their assistance.
\newpage

\printbibliography
\clearpage
\onecolumn

\renewcommand{\thefigure}{A\arabic{figure}}
\setcounter{figure}{0}
\section*{Appendix A: Additional samples and elemental maps}
\begin{figure*}[h]
    \centering
    \includegraphics[width=0.8\linewidth]{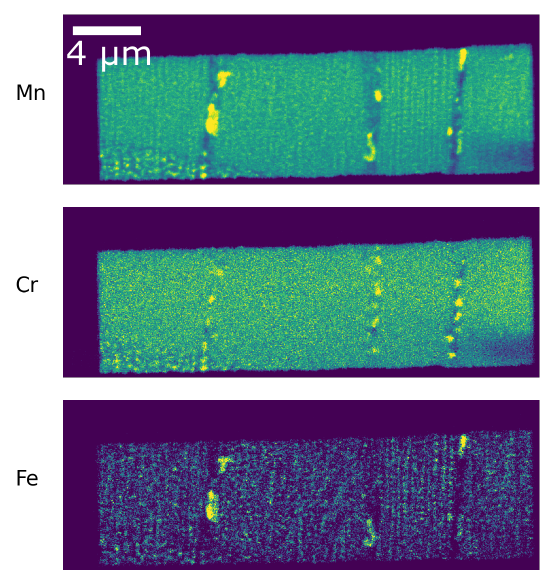}
    \caption{Ex-situ annealed sample. The ex-situ sample was directly heat-treated in bulk at 375 °C for 8 h, after which it was extracted and prepared by FIB-SEM.}
    \label{figA1}
\end{figure*}

\begin{figure*}
    \centering
    \includegraphics[width=\linewidth]{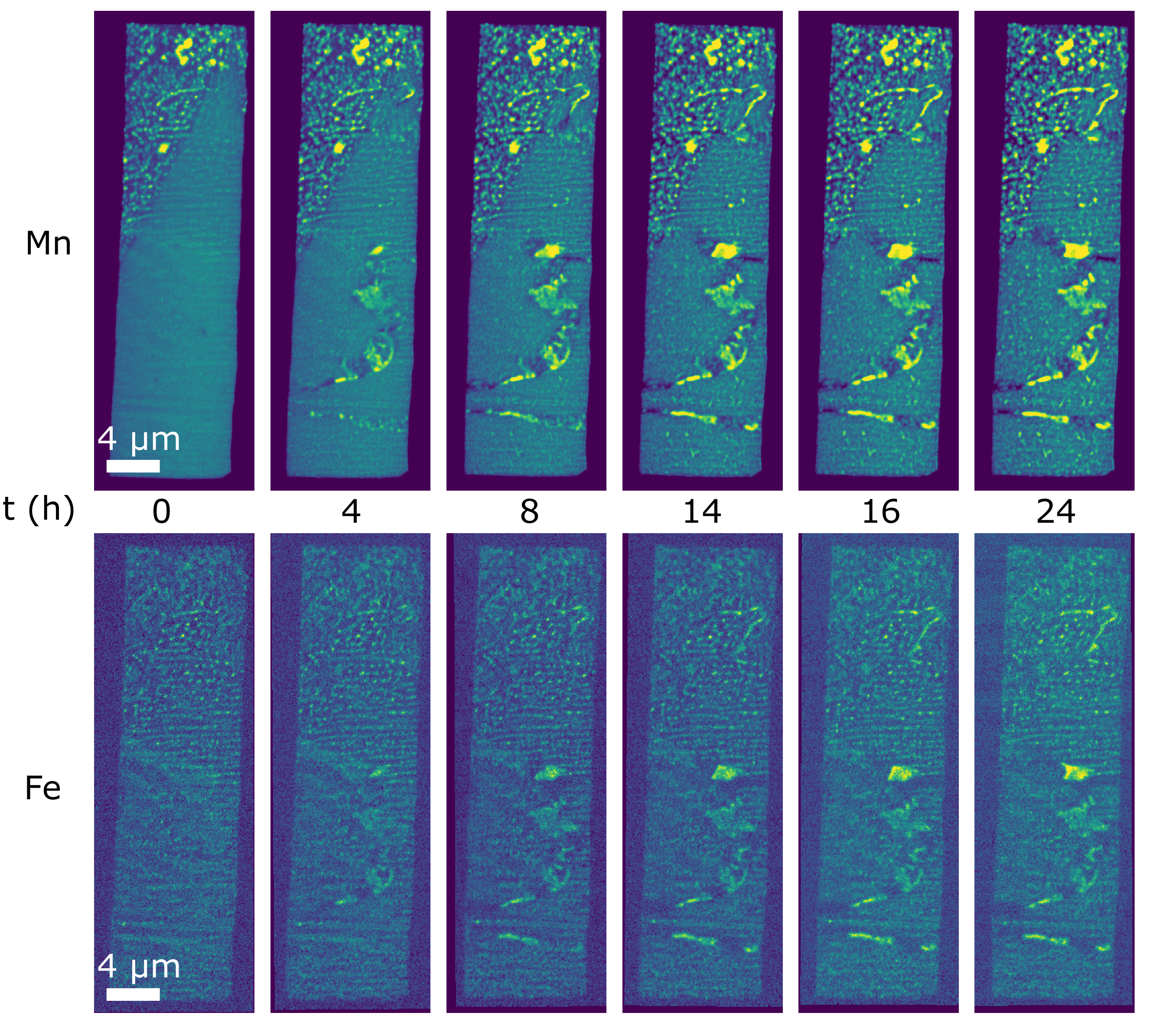}
    \caption{Elemental mapping of Mn and Fe showing Fe entering into the Mn-rich precipitates. Fe is not present in the Cr-rich precipitates nor the as-printed melt-pool boundary precipitates}
    \label{figA2}
\end{figure*}

Figure \ref{figA1} shows the distribution of Mn, Cr and Fe in the ex-situ annealed control sample. A lamella was extracted from a bulk sample that had been subject to a direct heat treatment at 375 °C for 8 hours. The microstructure looks qualitatively similar to the in situ results at the same time snapshot, with grain boundary precipitates grown up to 2 µm wide, and a clear distinction can be seen between Mn-rich and Cr-rich precipitates. Figure \ref{figA2} shows the distribution of Fe, along with Mn as a reference. It is evident that Fe dissolves into the larger Mn-rich precipitates at grain boundaries, but not in the Cr-rich phase. Fe is not consistently present in Mn-Cr-rich precipitates in the melt-pool boundary either.

\end{document}